\documentclass[aps,prd,amsmath,amssymb,preprintnumbers,nofootinbib,a4paper,11pt]{revtex4-1}
\pdfoutput=1
\usepackage{amsthm}
\usepackage{graphicx}
	\graphicspath{{./NewFig/}}
\usepackage{url}
\usepackage[bookmarks, pagebackref=false]{hyperref}
\usepackage{color}
	\definecolor{rossoCP3}{cmyk}{0,.88,.77,.40}
		\definecolor{graa}{rgb}{0.8,0.8,0.8}
		\definecolor{blaa}{rgb}{0.2,0.2,0.6}
		\hypersetup{
			colorlinks,
			bookmarksopen,
			bookmarksnumbered,
			citecolor=blaa, 		
			linkcolor=rossoCP3,	
			urlcolor=rossoCP3,			
			}
\usepackage{dcolumn}
\usepackage{bm}
\usepackage{bbm}
\usepackage{pxfonts}

\usepackage{epsfig}
\usepackage{placeins}

\usepackage[ margin=5pt, font=small,labelfont=bf, justification=raggedright]{caption}
\usepackage{slashed}
\usepackage{subfig}

\newcommand{\beq}{\begin{eqnarray}}
\newcommand{\eeq}{\end{eqnarray}}

\newcommand{\bmp}{\noindent\begin{minipage}{16cm}}
\newcommand{\emp}{\end{minipage}\vskip 7mm} 

\def\lsim{\mathrel{\rlap{\lower4pt\hbox{\hskip1pt$\sim$}}
    \raise1pt\hbox{$<$}}}                
\def\gsim{\mathrel{\rlap{\lower4pt\hbox{\hskip1pt$\sim$}}
    \raise1pt\hbox{$>$}}}                

\preprint{CP?-Origins-2013-32 DNRF90 and DIAS-2013-32}

\begin{document}

\title{\Large  \color{rossoCP3} The Technicolor Higgs in the light of LHC data}
\author{Alexander Belyaev$^{\color{rossoCP3}{\clubsuit\vardiamondsuit}}$}
\author{Matthew S. Brown$^{\color{rossoCP3}{\clubsuit}}$}
\author{Roshan Foadi$^{\color{rossoCP3}{\spadesuit}}$}
\author{Mads T. Frandsen$^{\color{rossoCP3}{\varheartsuit}}$}

\affiliation{ $^{\color{rossoCP3}{\clubsuit}}$ {School of Physics \& Astronomy, University of Southampton, 
Highfield, Southampton SO17 1BJ, UK}}

\affiliation{ $^{\color{rossoCP3}{\vardiamondsuit}}$ {Particle Physics Department, Rutherford Appleton Laboratory, 
Chilton, Didcot, Oxon OX11 0QX, UK}}

\affiliation{$^{\color{rossoCP3}{\spadesuit}}$ {Centre for Cosmology, Particle Physics and Phenomenology (CP3)
Chemin du Cyclotron 2 \\  Universit\'e catholique de Louvain, Belgium}}
 \affiliation{
$^{\color{rossoCP3}{\varheartsuit}}${ \color{rossoCP3}  \rm CP}$^{\color{rossoCP3} \bf 3}${\color{rossoCP3}\rm-Origins} \& the Danish Institute for Advanced Study {\color{rossoCP3} \rm DIAS},\\
University of Southern Denmark, Campusvej 55, DK-5230 Odense M, Denmark.
}
\begin{abstract}
We consider scenarios in which the 125~GeV resonance observed at the Large Hadron Collider is a Technicolor (TC) isosinglet scalar, the TC Higgs. By comparison with quantum chromodynamics, we argue that the couplings of the TC Higgs to the massive weak bosons are very close to the Standard Model (SM) values. The couplings to photons and gluons are model-dependent, but close to the SM values in several TC theories. The couplings of the TC Higgs to SM fermions are due to interactions beyond TC, such as Extended Technicolor: if such interactions successfully generate mass for the SM fermions, we argue that the couplings of the latter to the TC Higgs are also SM-like.
\\
We suggest a generic parameterization of the TC Higgs interactions with SM particles that accommodates a large class of TC models, and we perform a fit of these parameters to the Higgs LHC data. 
The fit reveals regions of parameter space where the form factors are of order unity and consistent with data at the 95\% CL, in agreement with expectations in TC theories. This indicates that the discovered Higgs boson is consistent with the TC Higgs hypothesis for several TC theories.
 \\
[.1cm]
{\footnotesize  \it Preprint: CP$^3$-Origins-22013-32 \& DIAS-2013-32}
 \end{abstract}

\maketitle


\section{Introduction}
The ATLAS and CMS collaborations have discovered a new resonance with the approximate mass of 125~GeV~\cite{Aad:2012tfa,Chatrchyan:2012ufa}. The observed decays to Standard Model (SM) diboson pairs, 
$\gamma \gamma$, $Z Z^*$ and $WW^*$, are in rough agreement with those expected from the SM Higgs. We therefore assume that this state is a scalar, as suggested  in particular by the observed decay rates into $Z Z^*$ and $WW^*$, as well as their angular distributions~\cite{Frandsen:2012rj,Coleppa:2012eh,Eichten:2012qb,Freitas:2012kw}. The most up to date combination of Higgs search results are presented in \cite{ATLAS-CONF-2013-034,Chatrchyan:2013lba}.

If strong dynamics like Technicolor (TC)~\cite{Weinberg:1975gm,Susskind:1978ms} is the origin of electroweak symmetry breaking, the Higgs boson can be identified with the lightest scalar resonance of the theory, the TC Higgs. In quantum chromodynamics (QCD) the analogous lightest resonance (except for the three pions) is indeed the CP-even scalar $f_0(500)$, also known as the $\sigma$ meson. The latter, however, is heavy compared to the pion decay constant $f_{\pi}$, and broad: $\Gamma_\sigma / m_\sigma \sim 1$. By analogy, the TC Higgs could {\it naively} be expected to be heavy, of the order of 1 TeV, and broad. This, however, would only be certainly true if two conditions were {\it both} satisfied: the TC Higgs is not coupled to SM particles, and the TC theory has QCD-like dynamics.

The first condition is certainly not satisfied: the TC Higgs does couple to SM particles. In particular, it couples to the top quark -- and to the other SM fermions -- through Extended Technicolor (ETC) interactions. These interactions are necessarily introduced in order for the SM fermions to ``communicate'' with the TC vacuum and acquire mass \cite{Dimopoulos:1979es,Eichten:1979ah}. As a consequence, large radiative corrections from the top quark can bring the TC Higgs mass down from ${\cal O}(1)$ TeV to 125~GeV, assuming the top Yukawa coupling to be $O(1)$~\cite{Foadi:2012bb}. The second condition need not be true either. If the TC theory, unlike QCD, has {\it walking} dynamics, then it is expected to feature a lighter scalar resonance, compared to QCD-like TC \cite{Yamawaki:1985zg,Sannino:2004qp,Hong:2004td}. In extreme-walking scenarios, such a light scalar is sometimes termed a {\it techni-dilaton} \cite{Yamawaki:1985zg,Holdom:1986ub,Holdom:1987yu,Elander:2009pk,Appelquist:2010gy,Evans:2013vca}, \emph{i.e.} the pseudo-Goldstone boson of spontaneous scale-symmetry breaking in TC.\footnote{The existence of a dilaton-like object has only been rigorously established in perturbative models such as {\it e.g.} \cite{Antipin:2011aa}} Finally, some TC models feature more than one dynamical scale, which in turn allows for light scalars to be part of the spectrum  \cite{Delgado:2010bb,Foadi:2012ks}.

A 125~GeV TC Higgs would also be narrow. We recall that the QCD $\sigma$ meson is broad because of its decay to two pions. In TC, however, the latter become the longitudinal components of the $W$ and $Z$ boson, and the corresponding TC Higgs decays are off-shell.\footnote{An example of this exists in the QCD scalar sector: the $f_0(980)$ resonance is narrow compared to the $f_0(500)$ because the {\it would-be} dominant decay mode into $K\bar{K}$ is off-shell.} In this paper we further argue that the couplings of the TC Higgs to massive weak bosons and SM fermions are expected to be SM-like. We also argue that the loop-induced couplings to photons and gluons are close to SM values in several TC theories. This motivates us to explore  in more detail the compatibility of the TC Higgs hypothesis with the properties of the discovered Higgs boson at 125 GeV.

Aside from the TC Higgs, new strong dynamics imply the presence of additional new resonances, which eventually will allow us to distinguish TC scenarios from theories with a fundamental Higgs \cite{Zerwekh:2005wh,Belyaev:2008yj}. Some of the models we consider here are being vigorously investigated on the lattice, such as the $SU(2)_{\rm Adj}$ MWT model~\cite{Catterall:2007yx,DelDebbio:2010hu}, and the $SU(2)_{\rm 2S}$ MWT model~\cite{Fodor:2012ty,Fodor:2012ni,Sinclair:2012fa,DeGrand:2013uha}. For a recent review see \cite{Appelquist:2013sia}.

The rest of this paper is organized as follows. In Sec.~\ref{sec:Eff}, after establishing an effective Lagrangian for the Higgs interactions with the SM particles, we estimate the corresponding couplings for the case of the TC Higgs: by comparison with QCD, through symmetry arguments, and through model computations. This leaves us with unknown form-factors for the couplings to photons and gluons, which we argue to be ${\cal O}(1)$. In Sec.~\ref{sec:ProdDec} we derive simple formulas for Higgs production and decay in TC. In Sec.~\ref{sec:Stat} we outline the statistical procedure for our fit of couplings and form-factors. In Sec.~\ref{sec:Fit} we show the results of our fit for several TC theories. Finally in Sec.~\ref{sec:Concl} we offer our conclusions.
\section{Effective Lagrangian for the TC Higgs} \label{sec:Eff}
The leading-order interactions of any type of Higgs boson $H$ with the SM particles can be parameterized by the effective Lagrangian
\begin{eqnarray}
{\cal L}_H &=&
\frac{2 M_W^2\ c_W}{v} H W^-_\mu W^{+\mu}
+ \frac{2 M_Z^2\ c_Z}{v} H Z_\mu Z^\mu
-\sum_f \frac{M_f\ c_f}{v} H\overline{f}f
+\frac{g_{H\gamma\gamma}}{v} H F_{\mu\nu} F^{\mu\nu}
+\frac{g_{Hgg}}{v} H G^a_{\mu\nu} G^{a\mu\nu}
\label{eq:Lagrangian}
\end{eqnarray}
where $v$ is the Higgs VEV, and the sum is over all quark and lepton flavors $f$. In the SM one has
\begin{equation}
c_W^{\rm SM} = c_Z^{\rm SM} = 1 \ , \quad
c_f^{\rm SM} = 1 \ ,
\label{eq:SMtree}
\end{equation}
for all quark and lepton flavors, and
\begin{eqnarray}
g_{H\gamma\gamma}^{\rm SM} &=& \frac{\alpha}{8\pi}\left|
F_1(\tau_W)+\sum_{f} N_c^f\ Q_f^2\ F_{1/2}(\tau_f)\right|  \label{eq:Hgammagamma} \ , \\
g_{Hgg}^{\rm SM} &=& \frac{\alpha_s}{16\pi}\left|
\sum_{q} F_{1/2}(\tau_q)\right| \label{eq:Hgg} \ .
\end{eqnarray}
The sum in the $H\gamma\gamma$ vertex is over all quark and lepton flavors $f$, whereas the sum in the $H gg$ vertex is over all quark flavors $q$. Here $Q_f$ is the QED charge of the flavor $f$, and $N_c^f$ is a color multiplicity factor that is equal to 1 for leptons and 3 for quarks. The functions $F_1$ and $F_{1/2}$ are defined by
\begin{eqnarray}
F_{1/2}(\tau_f) &\equiv& -2\tau_f\left[1+(1-\tau_f)f(\tau_f)\right] \nonumber \\
F_1(\tau_W) &\equiv& 2+3\tau_W+3\tau_W\left(2-\tau_W\right)f(\tau_W)\ ,
\end{eqnarray}
where
\begin{equation}
f(\tau)\equiv\left\{
\begin{array}{lr}
\displaystyle{\arcsin^2\frac{1}{\sqrt{\tau}}} & \tau \geq 1 \\
\displaystyle{-\frac{1}{4}\left[\log\frac{1+\sqrt{1-\tau}}{1-\sqrt{1-\tau}}-i\pi\right]^2} & \tau\leq 1
\end{array}
\right. \ ,
\end{equation}
and
\begin{equation}
\tau_f\equiv\frac{4M_f^2}{M_H^2}\ , \quad \tau_W\equiv\frac{4M_W^2}{M_H^2} \ .
\end{equation}
These functions quickly approach the asymptotic values $F_{1/2}(\tau_f) \ \to -\frac{4}{3}$ for $\tau_f \to \infty$ and $F_1(\tau_W)\  \to 7$ for $\tau_W \to \infty$. The partial decay widths into photons and gluons are
\begin{eqnarray}
\Gamma_{\gamma\gamma}^{\rm SM} &=& \frac{1}{4\pi} (g_{H\gamma\gamma}^{\rm SM})^2 \frac{M_H^3}{v^2}=\frac{\alpha^2}{256\pi^2}\frac{M_H^3}{v^2}
\left|
F_1(\tau_W)+\sum_{f} N_c^f\ Q_f^2\ F_{1/2}(\tau_f)\right|^2
\\
\Gamma_{gg}^{\rm SM} &=& \frac{2}{\pi} (g_{H gg}^{\rm SM})^2 \frac{M_H^3}{v^2} = \frac{\alpha_s^2}{128\pi} \frac{M_H^3}{v^2}
\left|
\sum_{q} F_{1/2}(\tau_q)\right|
\ .
\end{eqnarray}
We shall now argue that the couplings of the TC Higgs may be close to the SM values of Eqs.~(\ref{eq:SMtree})~--~(\ref{eq:Hgg}) under reasonable assumptions.
%
%
%
%
\subsection{TC Higgs couplings to $WW$ and $ZZ$}
To avoid large contributions to the $\rho$ parameter, we only consider TC theories with custodial isospin symmetry. The leading-order interaction of the TC Higgs with the electroweak Goldstone bosons, {\it i.e.} the technipions eaten by the $W$ and $Z$ bosons, can be described by the following dimension 5 operator (in analogy to QCD as described below): 
\begin{equation}
\mathcal{L}_{H\Pi\Pi}= \frac{c_\Pi}{v} H \partial_\mu\Pi^a \partial^\mu\Pi^a \ , \quad a=1,2,3.
\end{equation}
Ward identities and custodial symmetry then imply that the $HWW$ and $HZZ$ interactions of the TC Higgs are like the ones in Eq.~(\ref{eq:Lagrangian}), with
\begin{equation}
c_W = c_Z = c_\Pi \ .
\end{equation}
We expect $c_\Pi$ to be independent of $d(R_{\rm TC})$ and $N_{\rm TD}$, respectively the dimension of the technifermion representation under the TC gauge group, and the number of weak doublets carrying TC charge.\footnote{This is so because because $1/v$ scales like $[N_{\rm TD}\ d(R_{\rm TC})]^{-1/2}$ and this is also the scaling of a three-point function with external $\overline{Q}Q$ mesons in the large-$d(R_{\rm TC})$ and large-$N_{\rm TD}$ limits.}

We can estimate $c_\Pi$ in a TC theory with QCD-like dynamics by evaluating the corresponding quantity in two-flavor QCD: the $\sigma\pi\pi$ coupling  \cite{Harada:1995dc}. The relevant Lagrangian term is
\begin{equation}
\mathcal{L}_{\sigma \pi\pi} = \frac{c_\pi^{\rm QCD}}{f_\pi} \sigma \partial_\mu\pi^a \partial^\mu\pi^a\ , \quad a=1,2,3 ,
\label{Eq:qcdcoupl}
\end{equation}
where $f_\pi\simeq 93$ MeV is the pion decay constant (the QCD-scale equivalent of $v$ for a one-doublet TC model). A (complex) $\sigma\pi\pi$ coupling can be defined as the $I=0$ and $J=0$ partial-wave projection of the elastic $\pi\pi$ scattering amplitude at the $\sigma$ pole,  \emph{i.e.} \cite{GarciaMartin:2011jx}
\begin{eqnarray}
g_{\sigma\pi\pi}^2 \equiv - \lim_{s\to m_\sigma^2} 16\pi\ (s-m_\sigma^2)\
a_{00}(\pi\pi\to\pi\pi) \ ,
\label{eq:gspp}
\end{eqnarray}
where $m_\sigma$ is the complex mass of the $\sigma$ meson. Using Eq.~(\ref{Eq:qcdcoupl}), we see that the  $\sigma$ contributes to elastic $\pi\pi$ scattering with the invariant amplitude
\begin{eqnarray}
A(s,t,u) = -\left(\frac{c_\pi^{\rm QCD}}{f_\pi}\right)^2
\ \frac{(s-2m_\pi^2)^2}{s-m_\sigma^2}\ ,
\end{eqnarray}
where $s$, $t$ and $u$ are the usual Mandelstam variables.
The $I=0$ isospin projection is
\begin{eqnarray}
A_0 (s,x) = 3 A(s,t,u) + A(t,s,u) + A(u,t,s)\ ,
\end{eqnarray}
where $x$ is the cosine of the scattering angle. The $J=0$ parial wave projection of $A_0$ is
\begin{eqnarray}
a_{00}(s) = \frac{1}{64\pi}\sqrt{1-\frac{4m_\pi^2}{s}}\int_{-1}^1
dx\ A_0 (s,x)\ .
\end{eqnarray}
From the last four equations, we can derive the following relation
\begin{eqnarray}
\left|c_\pi^{\rm QCD}\right| = \sqrt{\frac{2}{3}} \
\frac{\left|g_{\sigma\pi\pi}\right| f_\pi}{\left|m_\sigma\right|^2}
\left|1-\frac{4 m_\pi^2}{m_\sigma^2}\right|^{-1/4}
\left|1-\frac{2 m_\pi^2}{m_\sigma^2}\right|^{-1}\ .
\label{eq:aQCD}
\end{eqnarray}
In Tab.~I we have summarised different fits \cite{Caprini:2005zr,Yndurain:2007qm,Oller:2003vf,Mennessier:2010ij,Pelaez:2010fj} of $m_\sigma$ and $\left|g_{\sigma\pi\pi}\right|$ reported in 
Ref.~\cite{GarciaMartin:2011jx}. These are shown in the first two columns of Tab.~\ref{tab:sigmapipi}, whereas the last column shows the value of $c_\pi^{\rm QCD}$ we computed using Eq.~(\ref{eq:aQCD}).\footnote{For simplicity, when the errors are asymmetric, we have taken the absolute value of the largest one to compute the error on $c_\pi^{\rm QCD}$.} All but one of the fits yield $\left|c_\pi^{\rm QCD}\right|\simeq 1$ with remarkable precision. This result suggests that the QCD $\pi^a-\sigma$ sector is well approximated by a linear sigma model, as the latter implies $c_\pi^{\rm QCD}=1$. 
This hypothesis is further reinforced by the fact that the $\pi NN$ coupling is approximately equal to the $\sigma NN$ coupling~\cite{Erkol:2005jz}. 
Therefore, if the TC dynamics is similar to that of QCD, we expect $c_\Pi\simeq 1$ and, thus, SM-like $HWW$ and $HZZ$ couplings. This need not be the case for walking dynamics, although departures from QCD-like dynamics by no means imply non-standard couplings to massive weak bosons. Finally we note that in TC the $HWW$ and $HZZ$ couplings receive small corrections from tree-level mixing of the electroweak bosons with the TC spin-one resonances\footnote{Similarly when vector mesons are introduced into the linear sigma model of QCD the fits are altered.} \cite{Belyaev:2008yj}.
\begin{table}[t!]
\centering
\begin{tabular}{|c|c|c|}
\hline
$m_\sigma$ (MeV) & $\left| g_{\sigma\pi\pi}\right|$ (GeV) & $\left|c_\pi^{\rm QCD}\right|$ \\
\hline \hline
$457_{-13}^{+14}-i(279_{-7}^{+11})$\ , \cite{GarciaMartin:2011jx} & $3.59_{-0.13}^{+0.11}$ & $1.0169\pm 0.06$ \\
\hline
$445\pm 25-i(278_{-18}^{+22})$\ , \cite{GarciaMartin:2011jx} & $3.4\pm 0.5$ & $1.0013\pm 0.17$ \\
\hline
$441_{-8}^{+16}-i(272_{-12.5}^{+9})$\, \cite{Caprini:2005zr} & $3.31_{-0.15}^{+0.35}$ & $1.0035 \pm 0.12$ \\
\hline
$474\pm 6-i(254\pm 4) $\ , \cite{Yndurain:2007qm} & $3.58\pm 0.03$ & $1.0264\pm 0.024$ \\
\hline
$443\pm 2-i(216\pm 4)$\, \cite{Oller:2003vf} & $2.97\pm 0.04$ & $1.0479\pm 0.020$ \\
\hline
$452\pm 12-i(260\pm 15)$ \, \cite{Mennessier:2010ij} & $2.65\pm 0.01$ & $0.8026\pm 0.053$ \\
\hline
$453-i 271$ \ , \cite{Pelaez:2010fj} & $3.5$ & $1.0255$ \\
\hline
\end{tabular}
\caption{Fits to $m_\sigma$ and $g_{\sigma\pi\pi}$ extrapolated from the elastic $\pi\pi$ scattering(see \cite{GarciaMartin:2011jx} for a summary of recent results). The last column gives $\left|c_\pi^{\rm QCD}\right|$ according to Eq.~(\ref{eq:aQCD}).}
\label{tab:sigmapipi}
\end{table}
\subsection{TC Higgs couplings to $ff$}
The interactions of the TC Higgs or the technipions with two SM fermions are due to four-fermion operators of the form $\bar{f} f^\prime \overline{F} F^\prime$. 
Here $f$ and $f^\prime$ are SM flavors, whereas $F$ and $F^\prime$ are techniflavors, with TC gauge indices contracted to form a TC singlet. 
At low energy these operators generate vertices such as  $H\bar{f}f$ and $\Pi^a\bar{f}f^\prime$.  Unlike the coupling to the weak bosons, there are no corresponding quantities in QCD which allow us to estimate the value of these couplings. However, Ward identities guarantee that the couplings of two SM fermions with the EW Goldstone bosons, at zero external momenta, are identical to their SM values. 
Of course these Ward identities tell us nothing about the $H\bar{f}f$ coupling. On the other hand, we have just seen that the interactions of the $\sigma$ meson and the pions in QCD are well approximated by a linear sigma model. If this holds in TC as well, then a $\Pi^a\bar{f}f^\prime$ coupling close to its SM value implies that the TC Higgs couplings to SM flavors are close to their SM values, {\it i.e} $c_f\simeq 1$. In the case of a true techni-dilaton, Ward-identities related to the spontaneously-broken scale symmetry can be used to determine $c_f$ \cite{Matsuzaki:2012vc,Matsuzaki:2012mk}. Finally, if SM fermion masses are generated by an ETC sector, we expect the TC Higgs couplings to SM fermions to be proportional to the SM fermion masses, and to be SM-like as argued in \cite{Foadi:2012bb}.
\subsection{TC Higgs couplings to $\gamma\gamma$ and $GG$}
\begin{figure}[t!]
\begin{center}
\includegraphics[width=10.0cm]{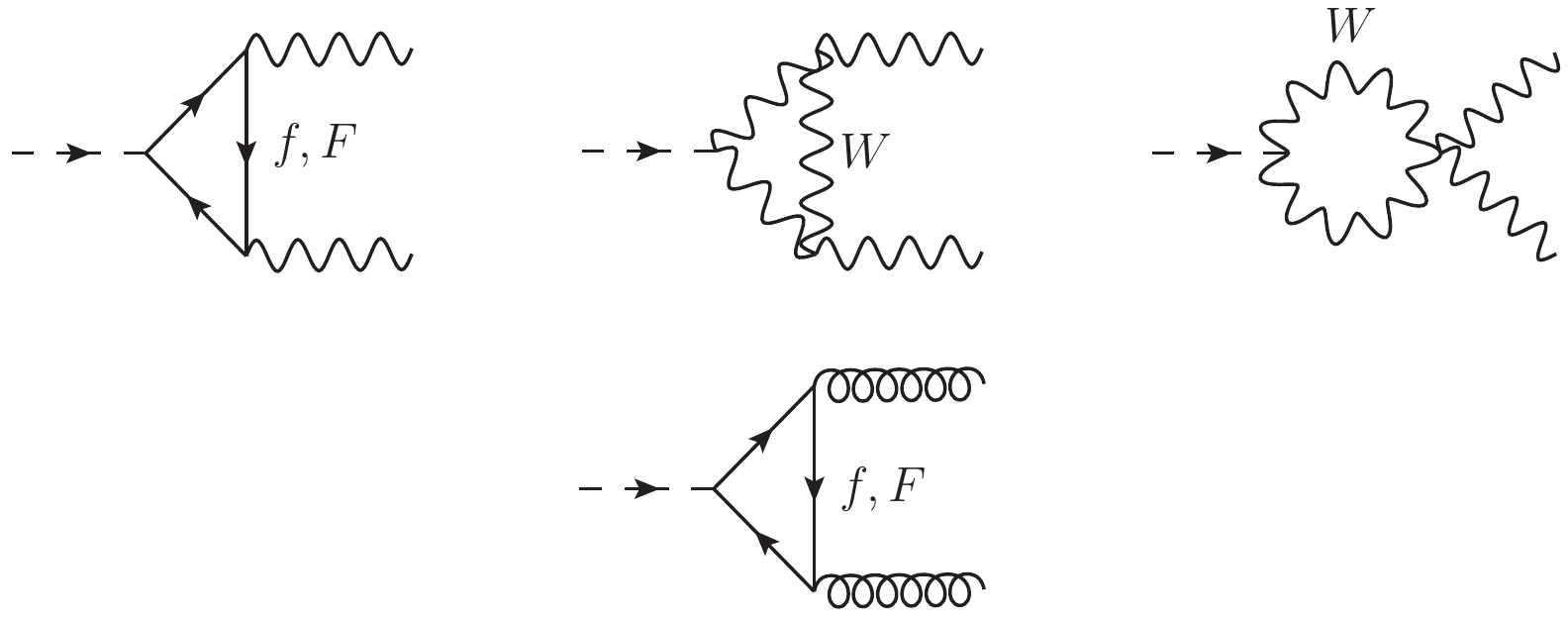}
\end{center}
\caption{One-loop amplitudes for Higgs decay into $\gamma\gamma$ and $gg$ pairs in TC.}
\label{fig:HiggsDecays}
\end{figure}
In order to compute the TC Higgs vertex with $\gamma\gamma$ and $GG$ we adopt a ``hybrid'' model \cite{Giacosa:2007bs,vanBeveren:2008st,Kalinovsky:2008iz,Volkov:2009mz}, in which the composite TC Higgs couples to the constituent techniflavors $F$ through a Yukawa-like vertex:
\begin{equation}
{\cal L}_{HFF} = - \sum_F \frac{M_F}{v} H\overline{F}F \ .
\end{equation}
Here the sum is over all TC Higgs constituent techniflavors. The dynamical technifermion mass $M_F$ can be estimated using the Pagels-Stokar relation,
\begin{equation}
v^2 = \frac{d(R_{\rm TC}) N_{\rm TD}}{4\pi^2}\ M_F^2\ \log\frac{\Lambda^2}{M_F^2} \ ,
\end{equation}
where $\Lambda\sim 4\pi F_\Pi=4\pi v/\sqrt{N_{\rm TD}}$. Unless $d(R_{\rm TC}) N_{\rm TD}$ is large, this gives $M_F$ of the order of hundreds of GeV. In practice we need not know the precise value of $M_F$, as our only dependence on it is through the loop function $F_{1/2}(\tau_F)$, which quickly reaches the asymptotic value for $\tau_F\equiv 4M_F^2/M_H^2>1$. 

The leading-order contributions to $H\to \gamma\gamma$ and $H\to gg$ are given by the diagrams of Fig.~\ref{fig:HiggsDecays}. Formally we are treating the TC Higgs as a pure $F\overline{F}$ bound state and the techniquark contributions to $H\to \gamma\gamma$ and $H\to gg$ should be regarded as large-$d(R_{\rm TC})$ results. Subdominant contributions, in $1/d(R_{\rm TC})$, arise from higher loop diagrams involving technipions and other resonances in this framework. These are model dependent, and we parametrize their effect by including overall form factors $a_{H\gamma\gamma}$ and $a_{Hgg}$, respectively, which approach unity in the large-$d(R_{\rm TC})$ limit. We note that in one of the TC theories we consider (see below) there are new heavy leptons carrying no TC charge: these contribute significantly to the $H\to \gamma\gamma$ process. Finally, in the $W$ loop we subtract 2 from $F_1(\tau_W)$, which corresponds to the contribution from the Goldstone bosons in Landau gauge \cite{Marciano:2011gm}. In fact the latter should be regarded as a contribution from the TC sector -- as the electroweak Goldstone bosons are technipions -- which is already included in the form factor $a_{H\gamma\gamma}$. The result for the effective couplings is
\begin{eqnarray}
g_{H\gamma\gamma}^{\rm TC} &=& \frac{\alpha}{8\pi}\left|
c_\Pi\left[F_1(\tau_W)-2\right]
+\sum_{f} c_f\ N_c^f\ Q_f^2\ F_{1/2}(\tau_f)
+a_{H\gamma\gamma}\ d(R_{\rm TC})
\sum_{F} \ N_c^F\ Q_F^2\ F_{1/2}(\tau_F)\right|  \ \label{eq:HgammagammaTC} ,  \\
g_{Hgg}^{\rm TC} &=& \frac{\alpha_s}{16\pi}\left|
\sum_{q} c_q^{\rm TC}\ F_{1/2}(\tau_q)
+ a_{Hgg}\  d(R_{\rm TC}) \sum_{F\in {\rm QCD}} F_{1/2}(\tau_F)\right| \label{eq:HggTC} \ ,
\label{Eq:loopcouplings}
\end{eqnarray}
where $N_c^f$ and $N_c^F$ are color multiplicity factors for the flavor $f$ and the techniflavor $F$, respectively, and the second sum in Eq.~(\ref{eq:HggTC}) is over colored techniflavors only.\footnote{Strictly speaking the dimension of the representation, $d(R_{\rm TC})$, should be inside the sums over $F$, as different techniflavors may belong to different representations. However in this paper we only consider TC theories in which the techniflavors carrying SM charges belong to a single representation.}

\begin{figure}[t!]
\begin{center}
\includegraphics[width=10.0cm]{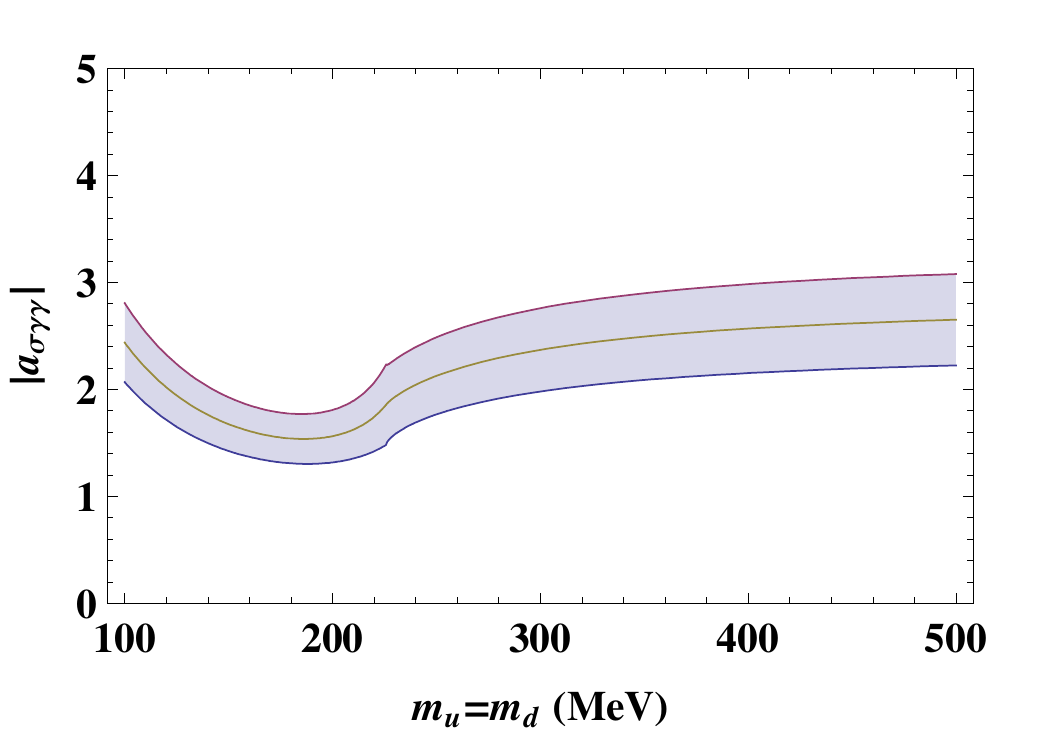}
\end{center}
\caption{Estimate of $\left|a_{\sigma\gamma\gamma}\right|\ (\pm 1\sigma)$ from the $\Gamma_{\sigma\to\gamma\gamma}$ partial width, as a function of the constituent $u$ and $d$ quark masses.}
\label{fig:bQCD}
\end{figure}

As mentioned above, the $a_{H\gamma\gamma}$ and $a_{Hgg}$ form-factors are model dependent. We can estimate the analogous form factor $|a_{\sigma\gamma\gamma}|$ in QCD by analyzing the decay of the $\sigma$ meson to two photons. Using the same hybrid model in which $\sigma$ couples to the constituent $u$ and $d$ quarks, the corresponding width is given by
\begin{eqnarray}
\Gamma_{\sigma\to\gamma\gamma} &=&
\frac{\alpha^2 \left({\rm Re}\ m_\sigma\right)^3 a_{\sigma\gamma\gamma}^2}{256\pi^3 f_\pi^2}
\left|3\ \left(\frac{2}{3}\right)^2 \
F_{1/2}\left(\frac{4 m_u^2}{\left({\rm Re}\ m_\sigma\right)^2}\right)
+ 3\ \left(-\frac{1}{3}\right)^2\
F_{1/2}\left(\frac{4 m_d^2}{\left({\rm Re}\ m_\sigma\right)^2}\right)
\right|^2 \ ,
\end{eqnarray}
where $m_u$ and $m_d$ are the {\em constituent} quark masses (not to be confused with the current quark masses). 
These depend on the model used to approximate low-energy QCD, and are expected to be around 300 MeV. 
The partial width $\Gamma_{\sigma\to\gamma\gamma}$ can be obtained from the latest Particle Data Group report~\cite{Eidelman:2004wy}. 
The average value is found to be $\Gamma_{\sigma\to\gamma\gamma}=2.79\pm 0.86$ keV. The average value of the real part of $m_\sigma$ given in Tab.~\ref{tab:sigmapipi} is ${\rm Re}(m_\sigma) = 451.7\pm 13.4\ {\rm MeV}$. The extracted value of  $\left|a_{\sigma\gamma\gamma}\right|\ (\pm 1\sigma)$ is shown in Fig.~{\ref{fig:bQCD}} as a function of the constituent quark mass. For $m_u=m_d=300$ MeV we obtain $\left|a_{\sigma\gamma\gamma}\right| = 2.37\pm 0.39$.
This result applies to QCD only, as the model-dependent contributions from pions and other resonances are important, although subdominant in a $1/N_c$ expansion.
We note that one may also attempt to compute the diphoton decay rate from an effective Lagrangian of composite states \cite{Bellazzini:2012tv,Carcamo-Hernandez:2013ypa,Castillo-Felisola:2013jua}.

\section{Higgs production and decay in TC}\label{sec:ProdDec}
Let $X$ denote the process (\emph{e.g.} gluon-gluon fusion) by which a Higgs boson is produced, and $Y$ denote its on-shell decay products. 
To a good approximation the number of events for the $XY$ process in a particular event category $c$ is
\begin{equation}
N_{XY}^c \equiv \sigma_{X}\times {\rm BR}_{Y}
\times \varepsilon_{XY}^c \times L\ ,
\end{equation}
where $\sigma_X$ is the $pp\to H$ production cross-section via the production process $X$, $\text{BR}_Y$ is the branching ratio of $H\to Y$, and $L$ is the integrated luminosity.
The efficiency factor $\varepsilon_{XY}^c$ (technically combining the cut acceptance and efficiency) gives the fraction of the total $XY$ events that are selected in event category $c$.
Currently there are 43 event categories between ATLAS and CMS, an example of which would be the ATLAS diphoton {\it unconverted, central, low-$p_T$} category.

If the cuts performed in event category $c$ were completely efficient at isolating one of the production processes (\emph{i.e.} so that all efficiency factors were zero except for one choice of $XY$) then one could usefully define a ``signal enhancement'' factor as the ratio between $N_{XY}^c$ and the corresponding SM prediction. 
This would give
\begin{equation}
\mu_{XY} = \widetilde{\sigma}_{X}\times \frac{\widetilde{\Gamma}_{Y}}{\widetilde{\Gamma}_\text{tot}} ,
\end{equation}
where a tilde denotes a (dimensionless) quantity expressed in SM units, \emph{e.g.} $ \widetilde{\sigma} \equiv \sigma/\sigma^{\rm SM}$.
In reality, no cut can be 100\% pure, and a category that is designed to isolate one particular method of Higgs production will invariably be contaminated by events from another production process.
Therefore, what is measured experimentally is the number of events inclusive of all production processes $X$. The signal enhancement factor is then
\begin{equation}
\mu_{Y}^c = \sum_X \widetilde{\sigma}_{X} R_{XY}^{c,\,\rm SM} \times \frac{\widetilde{\Gamma}_{Y}}{\widetilde{\Gamma}_\text{tot}}\ ,
\label{eq:muyc}
\end{equation}
where
\begin{equation}
    R_{XY}^{c,\,\rm SM} \equiv \frac{\sigma_{X}^{\text{SM}} \varepsilon_{XY}^c}{\sum_{X'} \sigma_{X'}^{\text{SM}} \varepsilon_{X'Y}^c}
    \label{eq:rfact}
\end{equation}
gives the fraction of Higgs bosons produced through the process $X$, in the SM, with acceptances and efficiencies for the final state $Y$ and event category $c$ included. 
As an example, if we consider again the ATLAS diphoton {\it unconverted, central, low-$p_T$} category, the fraction of the observed Higgs boson events produced through the gluon-gluon fusion process would be $93.7\%$~\cite{ATLAS-CONF-2013-012}, assuming the SM.
The total Higgs width in SM units is
\begin{eqnarray}
\widetilde{\Gamma}_\text{tot} &=& \sum_{f=b,c,\tau}\widetilde{\Gamma}_{ff} {\rm BR}_{ff}^{\rm SM}
+\sum_{V=W,Z,\gamma, g}  \widetilde{\Gamma}_{VV} {\rm BR}_{VV}^{\rm SM}
+ \widetilde{\Gamma}_{\text{else}} {\rm BR}_{\text{else}}^{\rm SM} \ ,
\label{eq:gamtot}
\end{eqnarray}
where ${\rm BR}_{\text{else}}^{\rm SM}\simeq 0.132\%$. Since the latter is a small fraction, instead of computing all remaining two- and multi-body decay channels, we shall simply take $\widetilde{\Gamma}_{\text{else}}=1$ and allow for little uncertainty in the final result.

\begin{figure}[t!]
\begin{center}
\includegraphics[width=15.0cm]{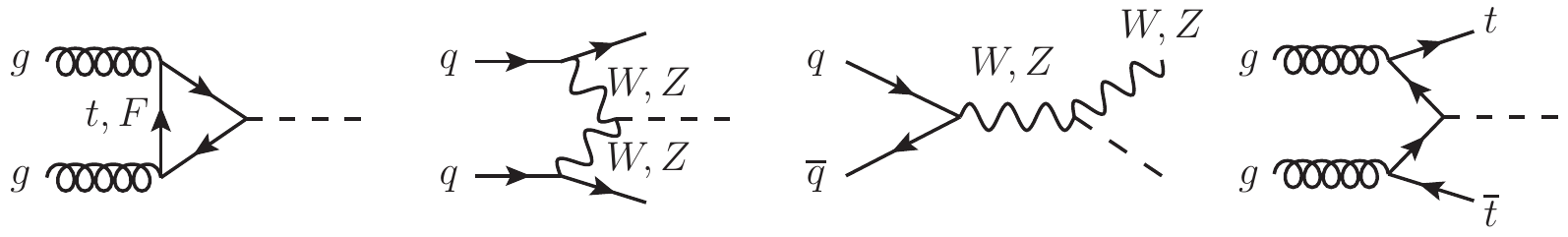}
\end{center}
\caption{Dominant Higgs production channels in TC.}
\label{fig:HiggsProduction}
\end{figure}

The dominant amplitudes for on-shell production of a TC-Higgs, at the LHC, are given by the diagrams shown in Fig.~\ref{fig:HiggsProduction}. The cross sections for gluon-gluon fusion, vector boson fusion, associated $W$ and $Z$ production and associated $t\bar{t}$ production, in SM units, are respectively
\begin{eqnarray}
&& \widetilde{\sigma}_{\text{ggF}} = \left(g_{Hgg}/g_{Hgg}^{\rm SM}\right)^2 \ , \\
&& \widetilde{\sigma}_{\text{VBF}} = c_\Pi^2 \ , \\
&& \widetilde{\sigma}_{WH} = \widetilde{\sigma}_{ZH}
= c_\Pi^2 \ ,  \\
&& \widetilde{\sigma}_{ttH} = c_t^2 \ ,
\end{eqnarray}
where $g_{Hgg}^{\rm SM}$ and $g_{Hgg}$ are computed as in Eqs.~(\ref{eq:Hgg}) and (\ref{eq:HggTC}), respectively. The relevant Higgs partial decay widths, expressed in SM units, are
\begin{eqnarray}
&&\widetilde{\Gamma}_{bb} = c_b^2 \ , \\
&&\widetilde{\Gamma}_{\tau\tau} = c_\tau^2 \ , \\
&&\widetilde{\Gamma}_{cc} = c_c^2 \ , \\
&& \widetilde{\Gamma}_{WW} = \widetilde{\Gamma}_{ZZ} 
= c_\Pi^2 \ , \\
&& \widetilde{\Gamma}_{gg} =
\left(g_{Hgg}/g_{Hgg}^{\rm SM}\right)^2 \ , \\ 
&& \widetilde{\Gamma}_{\gamma\gamma} 
= \left(g_{H\gamma\gamma}/g_{H\gamma\gamma}^{\rm SM}\right)^2 \ , 
\end{eqnarray}
where $g_{H\gamma\gamma}^{\rm SM}$ and $g_{H\gamma\gamma}$ are computed in Eqs.~(\ref{eq:Hgammagamma}) and (\ref{eq:HgammagammaTC}), respectively.
\section{Statistical procedure} \label{sec:Stat}
The normal method for confronting a beyond-the-Standard-Model theory with Higgs data (see, for example, \cite{Belanger:2013xza, Bechtle:2013xfa, Azatov:2012bz,Corbett:2012dm,Cacciapaglia:2012wb,Espinosa:2012im,Azatov:2012rd,Azatov:2012ga,Falkowski:2013dza,Giardino:2013bma})
is to make use of the best-fit values for the signal enhancement factors in each available final state $Y$ and cut category $c$. 
A $\chi^2$ test statistic
\begin{equation}
    \chi^2(\boldsymbol{\mu}) = \sum_{i, c, Y}  \frac{
    (\mu_Y^{c,i} - \hat{\mu}_Y^{c,i})^2}{\Delta(\mu_Y^{c,i})^2}
\end{equation}
is usually then formed, using the experimental collaborations' best-fit values of the enhancement factors ($\hat{\mu}$) and the given $1\sigma$ uncertainties $\Delta(\mu)$. The sum is taken over all final states $Y$, cut categories $c$ and experimental collaborations $i\in \{\text{ATLAS, CMS}\}$. 
One assumes this statistic follows a $\chi^2$ distribution with a number of degrees of freedom equal to the number of terms in the sum, and uses this to deduce the $p$-value for a particular hypothesis.

The difficulty with this method is in computing $\mu_Y^{c}$ in a particular model. 
One must evaluate the expression in Eq.~(\ref{eq:muyc}), but the experimental collaborations only make the efficiency factors (or, equivalently the ratios defined in Eq.~\eqref{eq:rfact}) available in the diphoton and $\tau^+\tau^-$ channels, so assumptions must be made for the other channels.
Another drawback of the procedure is that it neglects correlations between the systematic errors.

\begin{figure}[t!]
    \begin{center}
        \includegraphics{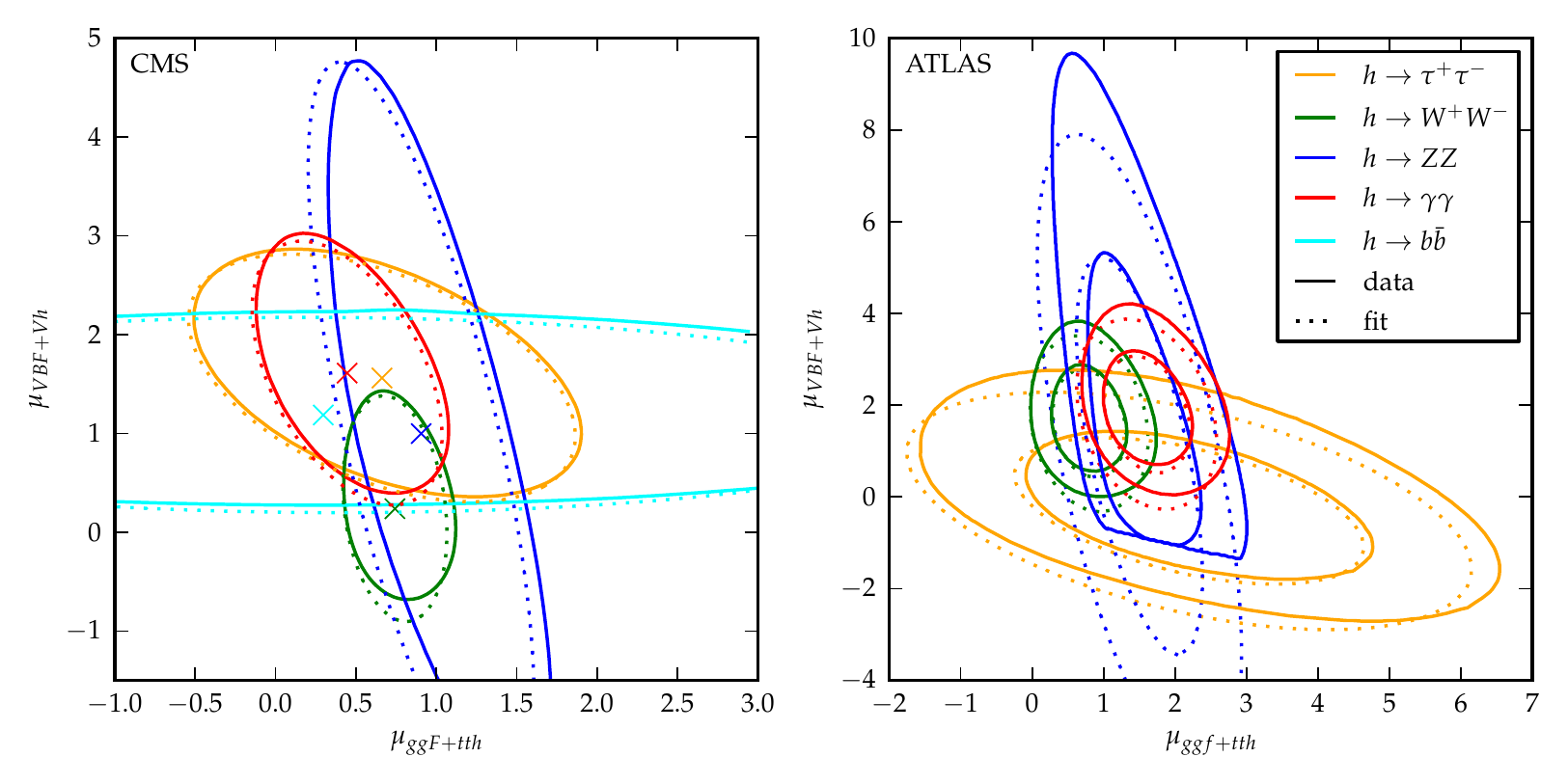}
        \caption{68\% CL (and 95\% CL in the ATLAS case) contours, comparing our fit (dotted lines) to official ATLAS and CMS fits.}
        \label{fig:fitdemo}
    \end{center}
\end{figure}

To alleviate the first problem, and ameliorate the second, we adopt a method, used for example in Refs.~\cite{Cacciapaglia:2012wb, Belanger:2013xza}, that makes use of the \emph{two-parameter} fits ATLAS and CMS have performed for each Higgs decay mode.  
These are presented in Figure~2 of \cite{ATLAS-CONF-2013-034} and Figure~4 of \cite{CMS-PAS-HIG-13-005} as 68\% (and also 95\% in the case of ATLAS) confidence level regions in the two-dimensional parameter space.
We reproduce the contours as the solid lines in our Figure~\ref{fig:fitdemo} for reference. 
The sharp cutoff in the $H \to ZZ^*\to 4\ell$ contour is due to the restriction that the likelihood must be zero anywhere in the parameter space where the total number of expected signal+background events is negative.

The collaborations perform the two-parameter fits in the following way. For a particular decay channel $Y$, they postulate a model identical to the SM except for factors enhancing the production cross-sections. 
One factor, $\mu_{Yg}$ enhances both the gluon fusion (ggH) and associated top production (ttH) mechanisms uniformly and the other, $\mu_{YV}$, enhances the vector boson fusion (VBF) and associated vector boson (VH) production. 
Assuming identical enhancements of ggH and ttH processes may be justified by the comparatively small SM ttH cross-section,\footnote{Although in \emph{e.g.} the OFTC model, new colored fermions enhance the ggH production cross-section while not affecting the ttH cross-section.} while equating the VBH and VH enhancements is reasonable because custodial symmetry is preserved to good accuracy.\footnote{The presence of resonances could change this picture due to the different kinematics in the channels.}

The next step is to form a likelihood function $\mathcal{L}(\boldsymbol{\mu}_Y, \boldsymbol{\theta})$, \emph{i.e.} a probability density function for observing a particular set of data, given a particular value of $\boldsymbol{\mu}_Y = (\mu_{Yg}, \mu_{YV})$, and the various ``nuisance parameters'' $\boldsymbol{\theta}$ that account for the systematic errors. 
From this, the ``profiled log likelihood ratio'' test statistic
\begin{equation}
    q_{\boldsymbol{\mu}_Y} = -2 \ln \left(
    \frac{
    \mathcal{L}(\boldsymbol{\mu}_Y, \hat{\boldsymbol{\theta}}_{\boldsymbol{\mu}_Y})}
    {\mathcal{L}(\hat{\boldsymbol{\mu}}_Y, \boldsymbol{\hat\theta})}\right)
\end{equation}
is formed, where $\hat{\boldsymbol{\theta}}_{\boldsymbol{\mu}_Y}$ is the value of $\boldsymbol{\theta}$ that maximizes the likelihood for a particular fixed $\boldsymbol{\mu}_Y$, and $\hat{\boldsymbol{\theta}}$ and $\hat{\boldsymbol{\mu}}_Y$ are global maximum-likelihood values.

If we assume that the two-dimensional parameterization above is true for some point in the parameter space, Wilks's theorem \cite{Wilks:1938}, as discussed in~\cite{Cowan:2011}, can be used to show that $q_{\boldsymbol{\mu}_Y}$ is distributed as a $\chi^2$ distribution with two degrees of freedom. 
If the probability density function for this distribution is denoted by $f_{\chi_2^2}(q_{\mu_Y})$ (where the subscript 2 signifies the number of degrees of freedom) then the $p$-value for a particular choice of $\boldsymbol{\mu}_Y$ is then given by
\begin{equation}
    p = \int_{q_{\boldsymbol{\mu}_Y}^{\text{obs}}}^\infty
    {
        f_{\chi^2_2}(q_{\boldsymbol{\mu}_Y}) \,\mathrm{d}q_{\boldsymbol{\mu}_Y}}.
\label{eq:pval}
\end{equation}
The contour plots presented by ATLAS and CMS are effectively plots of $q_{\boldsymbol{\mu}_Y}^{\text{obs}}$ as a function of $\boldsymbol{\mu}_Y$. 
The best-fit point has $q_{\boldsymbol{\mu}_Y}^{\text{obs}} = 0$ and the points on the 68\% CL contour (corresponding to $p=0.32$) have $q_{\boldsymbol{\mu}_Y}^{\text{obs}} \approx 2.3$.

If the data upon which the fit are based are distributed as a multivariate Gaussian (and this is a good approximation for numbers of events greater than around 10) then the test statistic takes the familiar ``chi square'' form, allowing for correlations in the errors:
\begin{equation}
    q_{\boldsymbol{\mu}_Y} \approx (\boldsymbol{\mu}_Y - \hat{\boldsymbol{\mu}}_Y)^{\mathrm{T}} \boldsymbol{\sigma}^{-1}_Y (\boldsymbol{\mu}_Y - \hat{\boldsymbol{\mu}}_Y).
\end{equation}
Suppressing the $Y$ index, the covariance matrix is conventionally parameterized as
\begin{equation}
    \boldsymbol{\sigma} = 
    \begin{pmatrix}
        \Delta_g^2             & \rho \Delta_g \Delta_V \\
        \rho \Delta_g \Delta_V & \Delta_V^2
    \end{pmatrix},        
\end{equation}
where $\Delta_{g}$ and $\Delta_{V}$ are the standard deviations in the $\mu^{g}$ and $\mu^{V}$ parameters and $\rho$ is the correlation coefficient.

We assume this bivariate Gaussian form and tune the covariance matrix to fit the 68\% CL contours in Figure~\ref{fig:fitdemo}. 
Our reproductions of the 68\% CL (and, for ATLAS, 95\% CL) contours, shown with dashed lines, are in good agreement with the official contours.
We next combine the likelihood functions by simply multiplying, which corresponds to adding the test statistics.
This gives
\begin{equation}
    q_{\boldsymbol{\mu}} = \sum_{i, Y} {(\boldsymbol{\mu}_Y^i - \hat{\boldsymbol{\mu}}_Y^i)^{\mathrm{T}} \boldsymbol{\sigma}^{-1}_{Y,i} (\boldsymbol{\mu}_Y^i - \hat{\boldsymbol{\mu}}_Y^i)},
\label{eq:qcomb}
\end{equation}
where again the index $i\in\{\text{ATLAS, CMS}\}$. For ATLAS, $Y\in\{\gamma\gamma, W^-W^+, ZZ, \tau^- \tau^+\}$ and for CMS, $Y\in\{\gamma\gamma, W^-W^+, ZZ, \tau^- \tau^+, b\bar{b}\}$.
Each channel contributes two degrees of freedom, $\mu_{Yg}$ and $\mu_{YV}$, to the test statistic and, with four channels from ATLAS and five from CMS, $q_{\boldsymbol{\mu}}$ will obey a $\chi^2$ distribution with $N_{\text{DOF}}=18$ degrees of freedom. 

We now have all the information to calculate the $p$-value for a particular choice of $\boldsymbol{\mu}$: 
the probability density function for $q_{\boldsymbol{\mu}}$ is a $\chi^2$ distribution for 18 degrees of freedom  and the value of $q_{\boldsymbol{\mu}}^{\text{obs}}$ as a function of $\boldsymbol{\mu}$ is found by fitting the contours in Figure~\ref{fig:fitdemo}.

All of the models we consider in this paper obey custodial symmetry and so are contained within the 18-parameter model described above. 
A particular point in a model's parameter space can be compared with experiment by calculating the 18 $\mu$ parameters. 
Unless the models explicitly incorporate details about the experimental apparatus, it will predict the same enhancement factors in a particular channel for both ATLAS and CMS. 
In terms of the parameters defined in \eqref{eq:Lagrangian}, the predicted enhancement factors would evaluate to
\begin{align}
    \mu_Y^g &= \left(\frac{g_{Hgg}}{g_{Hgg}^{\text{SM}}}\right)^2\times c_Y^2 \times \widetilde{\Gamma}_\text{tot}\\
    \mu_Y^V &= c_{\Pi}^2 \times c_Y^2 \times \widetilde{\Gamma}_\text{tot}
\end{align}
for $Y\in\{W^+W^-, ZZ, b\bar{b}, \tau^+ \tau^- \}$ and
\begin{align}
    \mu_{\gamma\gamma}^g &= \left(\frac{g_{Hgg}}{g_{Hgg}^{\text{SM}}}\right)^2 \times \left(\frac{g_{H\gamma\gamma}}{g_{H\gamma\gamma}^{\text{SM}}}\right)^2 \times \widetilde{\Gamma}_\text{tot}\\
    \mu_Y^V &= c_{\Pi}^2 \times \left(\frac{g_{H\gamma\gamma}}{g_{H\gamma\gamma}^{\text{SM}}}\right)^2 \times \widetilde{\Gamma}_\text{tot}
\end{align}
for the diphoton channel. 
The loop-level gluon and photon couplings $g_{H\gamma\gamma}$ and $g_{Hgg}$ are defined in \eqref{eq:HgammagammaTC} and \eqref{eq:HggTC} and the total width in SM units $\widetilde{\Gamma}_\text{tot}$ is defined in \eqref{eq:gamtot}.

With these calculated, $q_{\boldsymbol{\mu}}$ can be found using \eqref{eq:qcomb};
if the value exceeds 28.8 (corresponding to a $p$-value of 0.05) then the point $\boldsymbol{\mu}$ can be excluded at the 95\% confidence level.
\subsection{Assumptions and approximations}
Here we briefly bring together and reiterate the various assumptions and approximations we made in the above section.
The method we use relies on the two-parameter fit being a good parameterization of the true physics. 
It certainly has enough flexibility to fit the existing data well, and most candidate models of new physics respect custodial symmetry, motivating identical VBF and $VH$ enhancements as discussed above. 
This assumption, \emph{i.e.} that the ``two parameters per channel, per experiment'' parameterization is good, is required for Wilks's theorem to be applicable, and so any conclusions derived from this method should be interpreted in this way.

Advantageously, this method requires no assumptions about efficiency factors because, in the two-parameter fits, the collaborations make available purely theoretical variables with the experimental details unfolded. 
However, we \emph{are} making the assumption that the cut \emph{acceptances} for our signals are identical to the SM values: 
this corresponds to the assumption that the differential cross-sections predicted by the new physics models have the same shape as in the SM -- note that this is not true if \emph{e.g.} a resonance is present in $VH$ production \cite{Zerwekh:2005wh,Belyaev:2008yj}.
Another advantage of the method is that it includes correlations between systematic errors when combining the gluon fusion and vector production processes in the test statistic. However, correlations are necessarily neglected when summing over the final state channels $Y$.

Once the LHC resumes taking data, statistical uncertainties will be reduced and so systematic errors and their correlations will become more important. It will become increasingly useful, to theorists performing statistical tests of physics beyond the Standard Model, for the experimental collaborations to release more details, such as full likelihood functions in electronic format~\cite{Boudjema:2013qla}.

\section{Fit to ATLAS and CMS data} \label{sec:Fit}
Within a given TC theory the free parameters are $c_\Pi$, $c_f$, $a_{Hgg}$, $a_{H\gamma\gamma}$, the constituent techniflavor masses $M_F$, and, when present, the new lepton masses. However, the dependence of the observables on both technifermion and the new lepton masses is through the function $F_{1/2}(\tau_\psi)$, which quickly approaches the value $-4/3$ for masses above $M_H$. The heavy mass dependence is therefore very weak, and we can just set $F_{1/2}\to -4/3$ for these fermions. As a consequence our fit will only be on the parameters $c_\Pi$, $c_f$, $a_{Hgg}$ and $a_{H\gamma\gamma}$. Given current data we ignore $c_f$ for all flavors except for the top quark, the $b$-quark and the $\tau$-lepton. Thus the relevant six parameters to be fitted are $c_\Pi$, $c_b$, $c_t$, $c_\tau$, $a_{Hgg}$ and $a_{H\gamma\gamma}$. 

In the large-$d(R_{\rm TC})$ approximation, the form factors $a_{H\gamma\gamma}$ and $a_{Hgg}$ are equal to one. However, this is not necessarily an accurate description of TC for small values of $d(R_{\rm TC})$. Indeed, to fit data on the QCD $\sigma$-meson we needed a form factor $|a_{\sigma\gamma\gamma}|>1$. Therefore, in our fits, we also allow for arbitrary form factors. 
When fitting to Higgs data, only models with QCD-colored technifermions will be sensitive to $a_{Hgg}$, and one-family Technicolor (OFTC) is the only such model we consider in this paper. 
For all the other models, the fit is then effectively over five parameters: $c_\Pi$, $c_b$, $c_\tau$, $c_t$ and $a_{H\gamma\gamma}$.

Some of the theories we consider, like OFTC, have a rather large {\it naive}\footnote{By the naive $S$ parameter, we mean the contribution to the relevant vacuum polarization diagram from a one-loop computation with (dynamically) heavy technifermions $M_F\gg M_Z$.} $S$ parameter, whereas the minimal walking Technicolor theories (MWT) are constructed to minimize $S$. It should be noted, however, that negative contributions to $S$ may arise from different sources. For instance, the fact that the dynamical mass of the TC Higgs is much larger than the physical mass introduces a negative contribution to $S$. Also, isospin mass splitting between constituent technifermion masses (induced by ETC or the electroweak interactions) may reduce the value of $S$ and increase $T$, possibly leading to good agreement with electroweak precision data \cite{Appelquist:1993gi}. We will not address this issue here, leaving the analysis for a later work.
\subsection{General results}
The best-fit values we find are
\begin{eqnarray}
&& |c_\Pi| = 1.05030 \ , \  |c_b| = 1.08747 \ , \  |c_\tau| = 1.03835 \ , \nonumber \\
&& \left|g_{H\gamma\gamma}/g_{H\gamma\gamma}^{\rm SM}\right| = 1.17921 \ , \ \left|g_{Hgg}/g_{Hgg}^{\rm SM}\right| = 0.92234 \ .
\end{eqnarray}
From these we can obtain the best-fit values for $c_t$, $a_{H\gamma\gamma}$ and $a_{Hgg}$ in each TC theory (the best-fit values of $|c_\Pi|$, $|c_b|$ and $|c_\tau|$ are the same for each model). The $\chi^2$ value for the best-fit points is $\chi^2=11.75$. There are 18 degrees of freedom: five CMS channels plus four ATLAS channels, each with ggH and VBF values.
\subsection{Minimal walking Technicolor}
MWT is a class of theories in which the technifermion content consists of one weak technidoublet, $(U,D)$, and the dynamics is near-conformal (hence ``walking''). The two models defined in~\cite{Sannino:2004qp} are the $SU(2)_{\rm Adj}$ MWT model with one weak technidoublet in the adjoint representation, and the $SU(3)_{\rm 2S}$ MWT model with one weak technidoublet in the two-index symmetric (2S) representation (also known as ``next-to minimal walking Technicolor'', NMWT). Here we also consider $SU(3)_{\rm Adj}$ MWT with one weak technidoublet in the adjoint representation.
For even $N_{\rm TC}$ and the adjoint representation, a chiral lepton doublet, carrying no TC charge, is included in order to cure the topological Witten anomaly \cite{Witten:1982fp}. The fermion content is shown in Table~\ref{tab:TCtheories} below.
\begin{table}[t!]
\centering
\begin{tabular}{|l|c|c|c|c|c|c|c|}
\hline
TC theory & $F$ & $R_{\rm TC}$ & $Q$ & $N_c^F$ & $d(R_{\rm TC})
\sum_{F} N_c^F\ Q_F^2\ F_{1/2}(\tau_F)$ & $S_{\rm naive}$ \\
\hline \hline
$SU(2)_{F}$ MWT & $U$ & $\mathbf{2}$ & $1/2$ & 1 & $\simeq \frac{4}{3}$& $\frac{1}{3\pi } $ \\ \cline{2-5}
(UMT) \ & $D$ & $\mathbf{2}$ & $-1/2$ & 1 & & \\
\hline
$SU(2)_{Adj}$ MWT & $U$ & $\mathbf{3}$ & $(y+1)/2$ & 1 &$\simeq 4+20y^2$ & $\frac{1}{2\pi } $ \\ \cline{2-5}
\ & $D$ & $\mathbf{3}$ & $(y-1)/2$ & 1   & & \\ \cline{2-5}
\ & $N$ & $\mathbf{1}$ & $(-3y+1)/2$ & 1 & &\\ \cline{2-5}
\ & $E$ & $\mathbf{1}$ & $(-3y-1)/2$ & 1 & &\\
\hline
$SU(3)_{2S}$ MWT & $U$ & $\mathbf{6}$ & $1/2$ & 1 & $\simeq 4$ & $\frac{1}{\pi } $ \\ \cline{2-5}
(NMWT) \ & $D$ & $\mathbf{6}$ & $-1/2$ & 1 & & \\
\hline
$SU(3)_{Adj}$ MWT & $U$ & $\mathbf{8}$ & $1/2$ & 1 & $\simeq \frac{16}{3} $ & $\frac{4}{3\pi } $ \\ \cline{2-5}
\ & $D$ & $\mathbf{8}$ & $-1/2$ & 1 & & \\
\hline
WSTC (1D)/PGTC & $U_i$ & $\mathbf{N_{\rm TC}}$ & $1/2$ & 1 & $\simeq \frac{2}{3}N_{TC}$ & $\frac{N_{\rm TC}}{6\pi }$  \\ 
\cline{2-5}
\ & $D_i$ & $\mathbf{N_{\rm TC}}$ & $-1/2$ & 1 & & \\
\hline
WSTC (2D)  & $U$ & $\mathbf{N_{\rm TC}}$ & $(y+1)/2$ & 1 &$\simeq \frac{4}{3}N_{TC}(1+y^2)$ & $\frac{N_{\rm TC}}{3\pi }$ \\ 
\cline{2-5}
\ & $D$ & $\mathbf{N_{\rm TC}}$ & $(y-1)/2$ & 1   & & \\ \cline{2-5}
\ & $C$ & $\mathbf{N_{\rm TC}}$ & $-(y-1)/2$ & 1 & & \\ \cline{2-5}
\ & $S$ & $\mathbf{N_{\rm TC}}$ & $-(y+1)/2$ & 1 & & \\
\hline
OFTC  & $U$ & $\mathbf{N_{\rm TC}}$ & $(y+1)/2$ & 3 &$\simeq\frac{8}{3}N_{TC}(1+3y^2)$ & $\frac{2 N_{\rm TC}}{3\pi } $ \\ 
\cline{2-5}
\ & $D$ & $\mathbf{N_{\rm TC}}$ & $(y-1)/2$ & 3   & & \\ \cline{2-5}
\ & $N$ & $\mathbf{N_{\rm TC}}$ & $(-3y+1)/2$ & 1 & & \\ \cline{2-5}
\ & $E$ & $\mathbf{N_{\rm TC}}$ & $(-3y-1)/2$ & 1 & & \\
\hline
\end{tabular}
\caption{Table of the TC theories that we analyze, showing techniflavors (second column); the representation of the TC gauge group under which the techniflavors transform (third column); electric charges (fourth column); color multiplicity (fifth column); loop factors for the TC contribution to $g_{H\gamma\gamma}$ (sixth column); and the value of the {\it naive} $S$ parameter (seventh column). Note that among these theories, only the OFTC model has new fermions carrying QCD color charge.}
\label{tab:TCtheories}
\end{table} 

\begin{figure}[h!]
\includegraphics{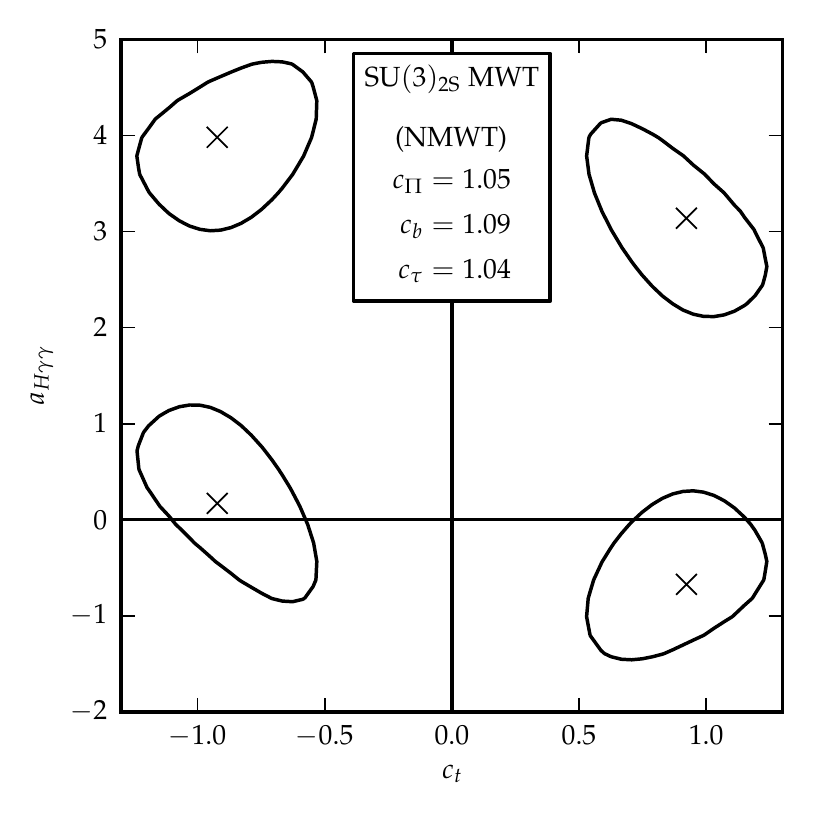}
\caption{$2\sigma$ exclusion contours for the $SU(3)_{\rm 2S}$ MWT model (NMWT model) in the plane of $c_t$ and $a_{H\gamma\gamma}$. All other parameters are fixed at their best-fit values shown in the legend.}
\label{fig:NMWT}
\end{figure}

\begin{figure}[h!]
\includegraphics{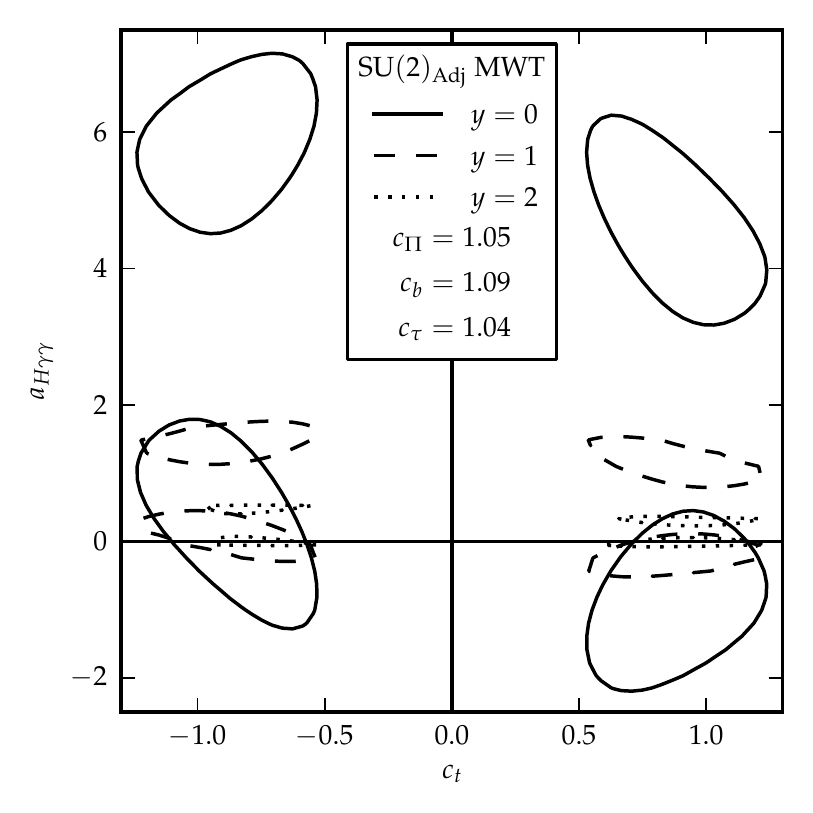}%
\includegraphics{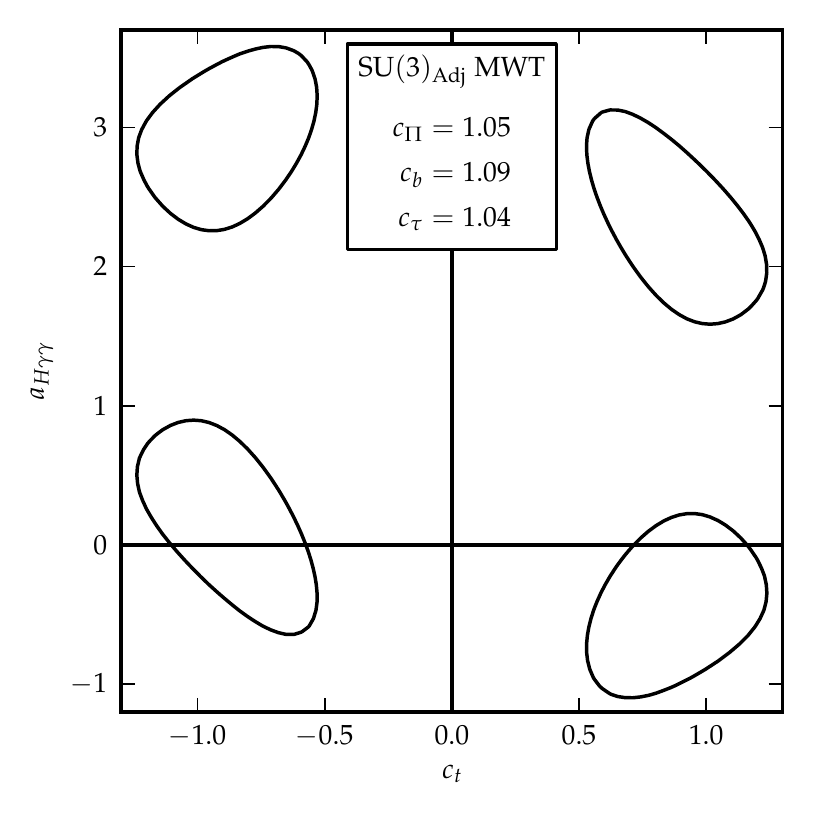}
\caption{$2\sigma$ exclusion contours for the $SU(2)_{\rm Adj}$ MWT model (left) and $SU(3)_{\rm Adj}$ MWT model (right). For the $SU(2)_{\rm Adj}$ MWT model we show three different hypercharge assignments: $y=0$ (solid), $y=1$ (dashed) and $y=2$ (dotted). All other parameters are fixed at their best-fit values shown in the legend.}
\label{fig:MWT}
\end{figure}

We present our fits to the LHC Higgs data for $SU(3)_{\rm 2S}$ MWT in Fig.~\ref{fig:NMWT}, and for $SU(2)_{\rm Adj}$ MWT and $SU(3)_{\rm Adj}$ MWT in Fig.~\ref{fig:MWT}. Note that while points inside the displayed contours in the $(c_t, a_{H\gamma\gamma})$ plane are allowed at the 95\% CL, we cannot conclude that points outside are not, as we keep $c_{\Pi}$, $c_b$ and $c_\tau$ fixed at their best fit values when drawing the contours. 
As usual in fits to LHC data, the contours display a degeneracy in the sign of the top Yukawa coupling. 
In general it is gratifying that we find values of the form factor norms $|a_{H\gamma\gamma}|$ similar to those extracted for the QCD $\sigma$ resonance in Fig.~\ref{fig:bQCD}. 

Consider first the $SU(3)_{\rm 2S}$ MWT model in Fig.~\ref{fig:NMWT}. From Table~\ref{tab:TCtheories} we find the relevant coupling for $H\to \gamma \gamma$  to be 
\begin{equation}
g_{H\gamma\gamma} \simeq \frac{\alpha}{8\pi} \left| 6-\frac{16}{9}c_t-4 a_{H\gamma\gamma}\right| 
\ , \quad 
g_{H\gamma\gamma}^{\rm SM}\simeq \frac{\alpha}{8\pi} \left|8-\frac{16}{9}\right|\simeq \frac{\alpha}{8\pi} 6 \ .
\label{eq:g_NMWT_approx}
\end{equation}
Here we have used approximate values for the loop functions, \emph{i.e.} $F_1(\tau_W) \simeq 8$ (subtracting 2 for the the longitudinal contribution in the left-hand expression), and $F_{1/2}(\tau_t) \simeq F_{1/2}(\tau_Q) \simeq -4/3$. We have also set $c_\Pi \simeq 1$. If $c_t\simeq 1$, a SM-like coupling of the TC Higgs to two photons is attained, in this model, for form factors $a_{H\gamma\gamma}\simeq -0.5$ and $a_{H\gamma\gamma}\simeq 2.5$. For $c_t\simeq -1$, Eq.~\eqref{eq:g_NMWT_approx} gives $a_{H\gamma\gamma}\simeq 0.4$ and $a_{H\gamma\gamma}\simeq 3.5$.  Indeed we see that all of these points sit inside the best fit contours in Fig.~\ref{fig:NMWT}. Intriguingly, $a_{H\gamma\gamma}\simeq 2.5$ is close to the $a_{\sigma\gamma\gamma}$ form factor in QCD, as shown in Fig.~\ref{fig:bQCD}. This is relevant, because the $SU(3)_{\rm 2S}$ MWT theory has the same global symmetries as two-flavor QCD, and we might expect the same type of contributions for the decay of the lightest scalar to two photons. Given that the TC Higgs arising in $SU(3)_{\rm 2S}$ MWT could also be as light as 125~GeV \cite{Foadi:2012bb,Fodor:2012ni,Fodor:2012ty}, this constitutes a possible candidate for the observed resonance at the LHC.

Next we consider the $SU(2)_{\rm Adj}$ (left) and $SU(3)_{\rm Adj}$ MWT (right) models in Fig.~\ref{fig:MWT}.
The $SU(3)_{\rm Adj}$ has more technifermions, by a factor 4/3, compared to the $SU(3)_{\rm 2S}$ MWT model and otherwise the same charge assignments.
 Accordingly for $c_t \simeq 1$ the norm of the best fit form factors are compressed towards slightly smaller values, and again the two contours centered near $c_t\simeq -1$ are shifted to slightly larger values of $a_{H\gamma\gamma}$.
Conversely, in the left panel of Fig.~\ref{fig:MWT} the $SU(2)_{\rm Adj}$ with $y=0$ has fewer technifermions, by a factor 1/2, and again identical charges.
 Although the heavy lepton doublet gives a slight contribution, the norms of the best fit form factors are dilated to larger values: note the change of scale in the two plots. As we increase the hypercharge parameter $y$ the contribution of the techniquarks and the heavy lepton doublet increases, and the contours are again compressed towards smaller norms of the form factors. This is shown by the dashed ($y=1$) and dotted ($y=2$) contours. Note that the global symmetries of the $SU(2)_{\rm Adj}$ and $SU(3)_{\rm Adj}$ MWT theories are different than those of two-flavor QCD, and we might therefore not expect the form factor to be QCD-like. In addition, the dynamics of these models are presumably very near-conformal if not conformal \cite{Catterall:2007yx,DelDebbio:2010hu}. We leave a study of the impact of this for future work. 

\subsection{Weinberg-Susskind, partially-gauged and two-scale Technicolor}
These classes of TC theories all feature a set of $N_{\rm TD}$ colorless weak technidoublets transforming in the fundamental representation of $SU(N_{\rm TC})$. In Weinberg-Susskind Technicolor (WSTC)~\cite{Weinberg:1975gm,Susskind:1978ms} these $N_{\rm TD}$ technifermions constitute the complete TC sector. Near-conformal dynamics is then achieved by taking a sufficiently large $N_{\rm TD}$. This, however, might lead to unacceptably large contributions to the $S$ parameter. In  partially-gauged Technicolor (PGTC)~\cite{Dietrich:2005jn,Luty:2008vs,Ryttov:2008xe} this problem is avoided by assigning electroweak charge to just one technidoublet. Near-conformality is then achieved by allowing for additional electroweak-neutral technifermions in the fundamental representation. Similarly, two-scale Technicolor theories (2STC)~\cite{Lane:1989ej} feature one weak technidoublet in the fundamental representation, and additional electroweak-neutral technifermions in higher-dimensional representations, once again included to achieve near-conformal dynamics. An example of 2STC is ultra-minimal Technicolor (UMT), an $SU(2)_{\rm TC}$ theory with one fundamental technidoublet and two electroweak-neutral technifermions in the adjoint representation. UMT is the TC theory with the smallest naive $S$ parameter while still featuring near-conformal dynamics. 

\begin{figure}[h!]
\includegraphics{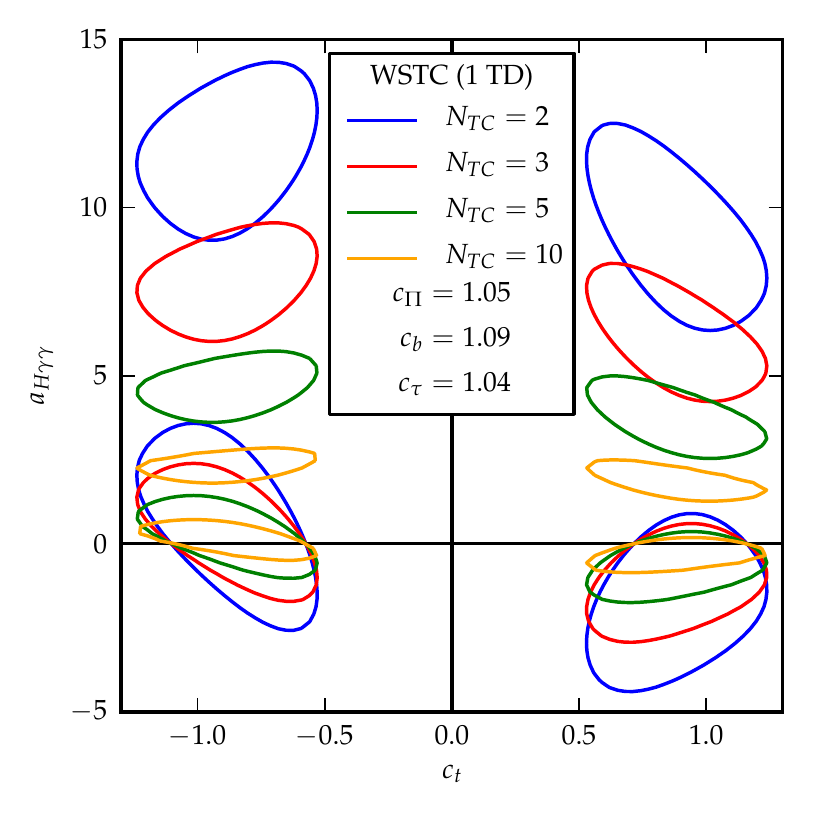}%
\includegraphics{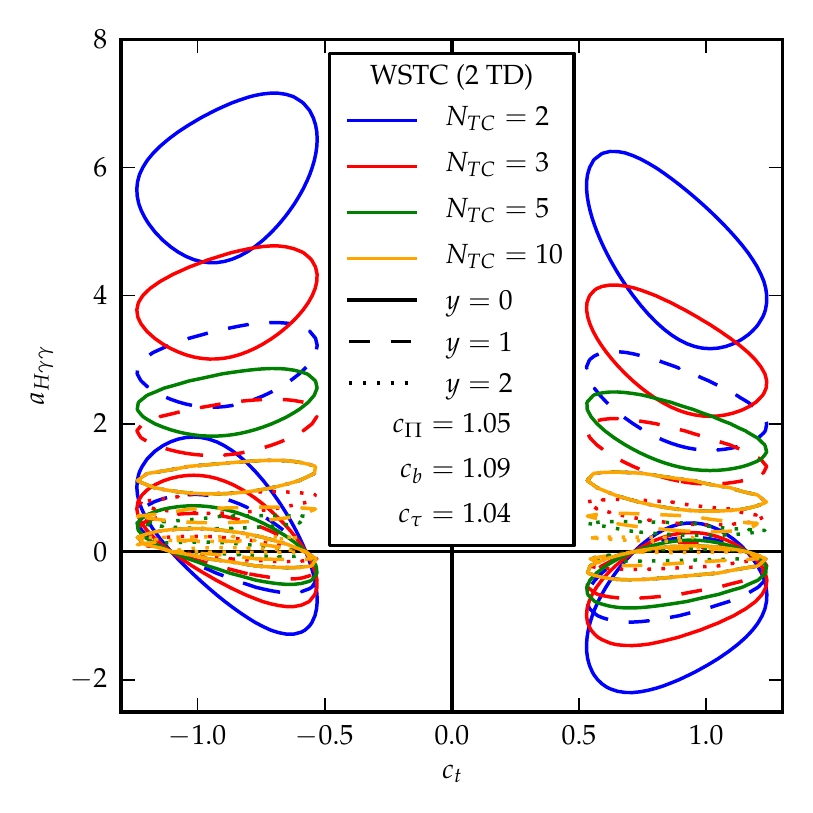}
\caption{Left: $2\sigma$ exclusion contours of WSTC with one weak technidoublet and different numbers of technicolors. These fits apply also to PGTC and 2STC. Right: WSTC with two weak technidoublets, varying numbers of technicolors, and different values of the hypercharge parameter $y$. The parameters not plotted are set to their best-fit values, given in the legends.}
\label{fig:WSTC}
\end{figure}

For our fits, PGTC and 2STC are equivalent to WSTC with one weak technidoublet, as long as the extra electroweak-neutral technifermions carry no color charge. See Table~\ref{tab:TCtheories} below for the matter content and the anomaly-free charges. We show the best fit values for $c_t$ and $a_{H\gamma\gamma}$ in Fig~\ref{fig:WSTC}. The contours follow the discussion above: the more technifermion degrees of freedom and/or the higher the charges of the fermions ($y> 0$), the smaller the norm of $a_{H\gamma\gamma}$ required to fit the data. It is worth stressing that non-QCD-like $a_{H\gamma\gamma}$ form factors might be expected in those theories, {\em e.g.} UMT, featuring different global symmetries than two-flavor QCD.

\subsection{One-family Technicolor}
Finally, we consider one-family Technicolor (OFTC) \cite{Farhi:1979zx}. This is a class of theories in which new generations of quarks and leptons are added to the SM and assigned TC charge. For $N_{\rm TC}=2$, based on a Schwinger-Dyson gap equation in the ladder approximation, near-conformal dynamics is expected to be achieved by including one full family. This expectation is currently being investigated on the lattice \cite{Bursa:2010xn,Ohki:2010sr,Hayakawa:2012gf}. The corresponding matter content is again shown in Table~\ref{tab:TCtheories}, from which we find the relevant couplings of $H\to \gamma \gamma$ and $H\to gg$ to be 
\begin{equation}
g_{H\gamma\gamma} \simeq \frac{\alpha}{8\pi} \left| 6-\frac{16}{9}c_t-\frac{8}{3} N_{\rm TC} (1+3y^2)\ a_{H\gamma\gamma}\right|  
\ , \quad 
g_{H g g} \simeq \frac{\alpha_s}{16\pi} \left| -\frac{4}{3}c_t - \frac{8}{3} N_{\rm TC}\ a_{Hgg} \right| \ .
\end{equation}
Here we have used approximate values for the loop functions and set $c_\Pi \simeq 1$, as in Eq.~\eqref{eq:g_NMWT_approx}. From these expressions it is clear that there is a degeneracy between the three quantities $a_{H\gamma\gamma}$, $c_t$ and $a_{Hgg}$ and the two observables $g_{H\gamma\gamma}$ and $g_{H g g}$. It also follows that for $N_{\rm TC}=2$ and $c_t\simeq 1$ we get a SM Higgs-like gluon fusion production rate for $a_{Hgg}\simeq -0.5, 0$.

\begin{figure}[!h]
    \includegraphics{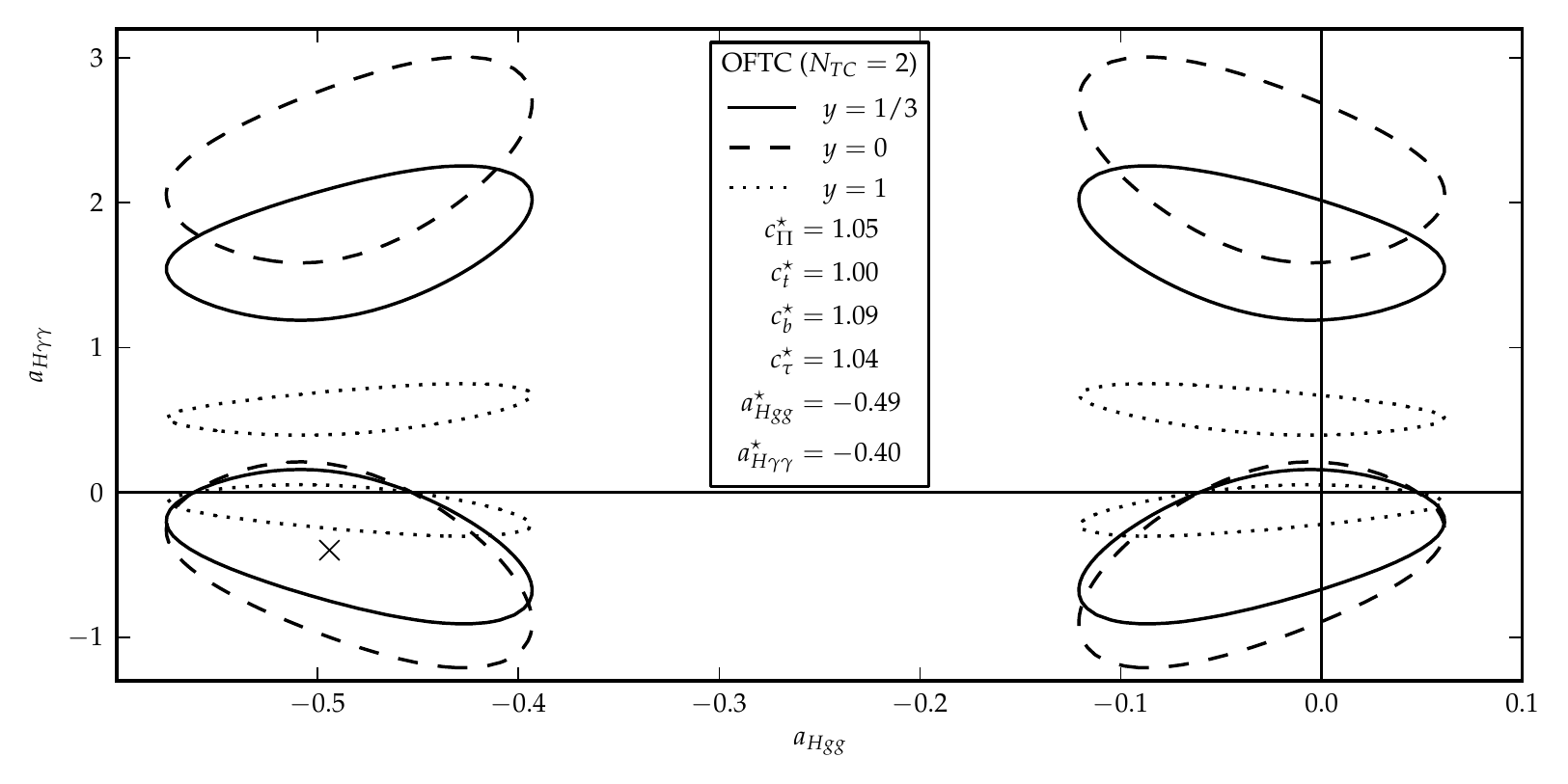}
    \includegraphics{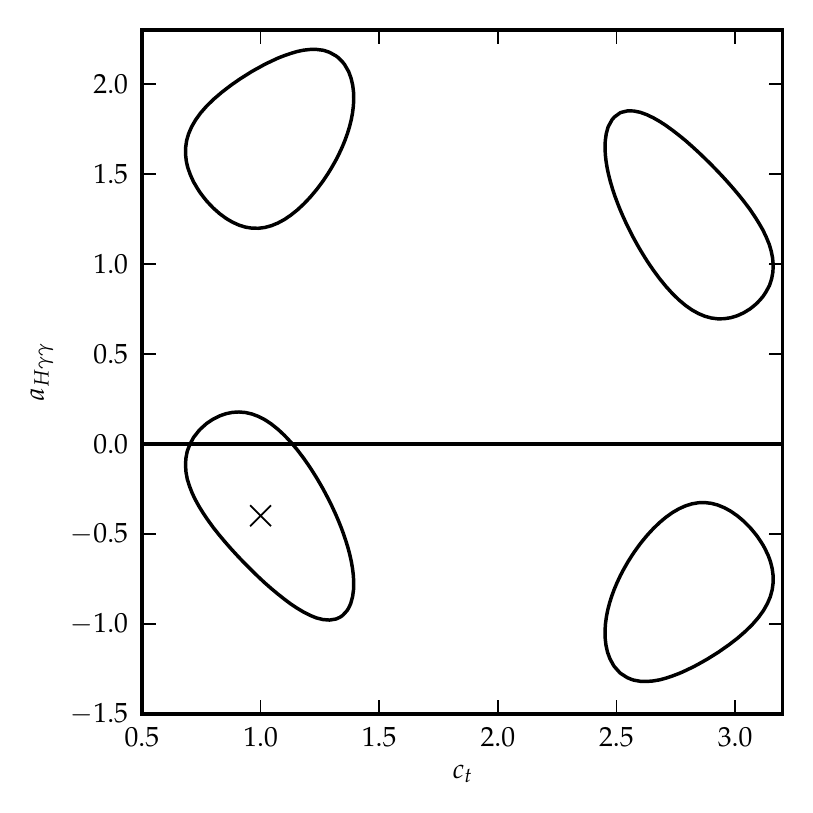}%
    \includegraphics{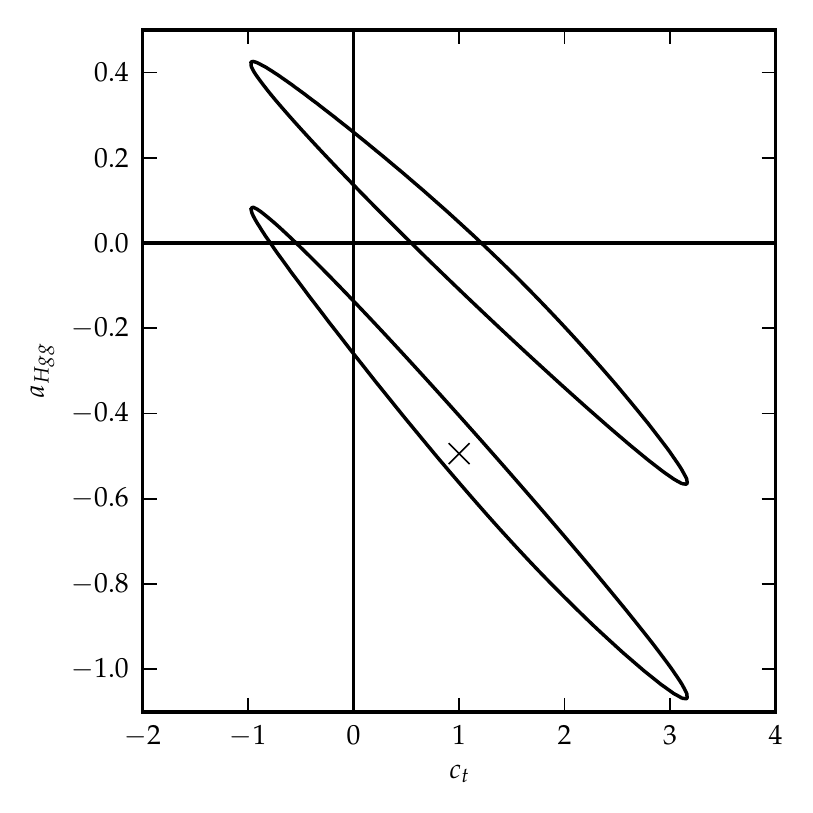}
    \caption{$2\sigma$ exclusion contours for one-family Technicolor (OFTC) with $N_{\rm TC}=2$ in the three planes of interest. 
    The first subfigure shows the contours for several values of the hypercharge parameter $y$, while the other plots fix $y=1/3$. 
    The stars denote the $y=1/3$ best fit coordinates.}
\label{Fig:OFTC}
\end{figure}

We show the best fit values for the parameters of the OFTC model in Fig~\ref{Fig:OFTC}. 
To break the degeneracy between $c_t$, $a_{Hgg}$ and $a_{H\gamma\gamma}$ we set $c_t=1$.
The top plot shows the best fit contours in the $a_{H\gamma\gamma},a_{Hgg}$ plane, for different values of the hypercharge parameter, $y=0,1/3,1$. All the remaining parameters are fixed: $c_t^\ast=1.0$, and the remaining ones at their best fit values. For $y=1/3$ these values are indicated by the starred quantities in the legend. As expected, $a_{Hgg}^\star \simeq -0.5$.  In general it is only possible to have form factors $|a_{Hgg}|\sim O(1)$ for $c_t$ significantly larger than 1. On the one hand this would imply a significantly enhanced $ttH$ production, providing a powerful experimental test. On the other hand, within typical ETC extensions of TC such a large $c_t$ is not easily accommodated.

In the bottom left panel we show the best fit contours of $a_{H\gamma\gamma}$ and $c_t$ as done for the other TC models above. Again we fix the hypercharge parameter at $y=1/3$ and keep all other parameters at their best fit values. The many technifermion degrees of freedom implies a relatively small value of $|a_{H\gamma\gamma}|$ to fit the data for $c_t\simeq 1$. Finally we illustrate the correlation between $a_{Hgg}$ and $c_t$ in the lower right panel of Fig~\ref{Fig:OFTC}, with the other parameters, including $a_{H\gamma\gamma}$, fixed to the best-fit values. Along the axis of the ellipses, the TC Higgs production via gluon fusion is close to that of the SM Higgs.  However once $c_t$ deviates too much from unity, the diphoton production rate is driven too far from the SM Higgs value to allow for a good fit to the data.

\section{Conclusions} \label{sec:Concl}
In this paper we have studied the Technicolor Higgs, \emph{i.e}. the lightest $J^{PC}=0^{++}$ resonance in various Technicolor models. It had previously been argued that this resonance can be as light as the observed 125 GeV boson \cite{Yamawaki:1985zg,Sannino:2004qp,Hong:2004td,Appelquist:2010gy,Foadi:2012bb}. Here we have further argued that the TC Higgs has SM-Higgs-like tree-level couplings. We then employed a simple model computation of the diphoton decay rate, including a form factor $a_{H\gamma\gamma}$ encoding strong coupling effects. In order to have an idea of the size of $a_{H\gamma\gamma}$, we computed the analogous quantity in QCD for the decay of the $\sigma$ meson into two photons, and obtained $a_{\sigma\gamma\gamma}\sim 2.5\gtrsim 1$: this is consistent with general expectations, as the form factor should approach unity in the large-$N_c$ limit. 

We have fitted several TC Higgs scenarios to LHC Higgs data. Our findings show TC Higgs couplings to massive weak bosons and SM fermions that are very close to SM values, in agreement with expectations for TC theories. In particular, the top quark Yukawa coupling is found to be close to unity: this is the right size needed if top corrections are to help provide a light mass for the TC Higgs~\cite{Foadi:2012bb}. The form factor for the diphoton channel is found to be $a_{H\gamma\gamma}\gtrsim 1$ for most of the TC theories we considered: as explained above, this is consistent with TC dynamics. We have also considered a theory with technifermions carrying QCD charge, namely one-family Technicolor. In this theory, in addition to the $a_{H\gamma\gamma}$ form factor, there is also a $a_{Hgg}$ form factor for TC Higgs decays into two gluons. Our analysis shows best fit values of these form factors that are less than one, a result which might be difficult to account for in TC.
 
We conclude that the TC Higgs is a viable candidate for the observed 125~GeV resonance at the LHC. For instance, a promising candidate TC theory is the $SU(3)_{2S}$ MWT (NMWT) model. In fact, this has the same global symmetries as two-flavor QCD, and the fit yields $a_{H\gamma\gamma}\sim a_{\sigma\gamma\gamma}$. Furthermore, the $SU(3)_{2S}$ MWT model has been investigated on the lattice, and the results are consistent with a 125~GeV TC Higgs~\cite{Foadi:2012bb,Fodor:2012ni,Fodor:2012ty}. There are several other TC theories consistent with current data. The ultimate verification of the TC scenario would of course be the discovery of new resonances.

\section*{Acknowledgments}
We thank Professor Glen Cowan for invaluable discussions regarding the statistical fit performed herein. The work of R.F. is supported by the Marie Curie IIF grant proposal 275012. M.T.F. acknowledges partial support from a `Sapere Aude' Grant no. 11-120829 from the Danish Council for Independent Research and partial support from the Danish National Research Foundation, grant number DNRF90. A.B. and M.S.B thank the NExT Institute for partial support.
\bibliography{tc_fit}

\end{document}